# Intrinsic Paramagnetic Meissner Effect due to s-wave Odd Frequency Superconductivity

A. Di Bernardo,[1] Z. Salman,[2] X. L. Wang,[3] M. Amado,[1] M. Egilmez,[4] M. G. Flokstra,[5] A. Suter,[2] S. L. Lee,[5] J. H. Zhao,[3] T. Prokscha,[2] E. Morenzoni,[2] M. G. Blamire,[1] J. Linder,[6] J. W. A. Robinson*[1]

[1]*Department of Materials Science and Metallurgy, University of Cambridge, 27 Charles Babbage Road, Cambridge CB3 0FS, United Kingdom.*
[2]*Laboratory for Muon Spectroscopy, Paul Scherrer Institute, 5232 Villigen PSI, Switzerland*
[3]*State Key Laboratory of Superlattices and Microstructures, Institute of Semiconductors, Chinese Academy of Sciences, Beijing 100083, China.*
[4]*Department of Physics, American University of Sharjah, Sharjah 26666, United Arab Emirates.*
[5]*School of Physics and Astronomy, SUPA, University of St. Andrews, St. Andrews KY16 9SS, United Kingdom.*
[6]*Department of Physics, Norwegian University of Science and Technology, N-7491 Trondheim, Norway.*

**In 1933, Meissner and Ochsenfeld reported the expulsion of magnetic flux – the diamagnetic Meissner effect – from the interior of superconducting lead. This discovery was crucial in formulating the Bardeen-Cooper-Schrieffer (BCS) theory of superconductivity. In exotic superconducting systems BCS theory does not strictly apply. A classical example is a superconductor-magnet hybrid system where magnetic ordering breaks time-reversal symmetry of the superconducting condensate and results in the stabilisation of an odd-frequency superconducting state. It has been predicted that under appropriate conditions, odd-frequency superconductivity should manifest in the Meissner state as fluctuations in the sign of the magnetic susceptibility meaning that the superconductivity can either repel (diamagnetic) or attract (paramagnetic) external magnetic flux. Here we report local probe measurements of faint magnetic fields in a Au/Ho/Nb trilayer system using low energy muons, where antiferromagnetic Ho (4.5 nm) breaks time-reversal symmetry of the proximity induced pair correlations in Au. From depth-resolved measurements below the superconducting transition of Nb we observe a local enhancement of the magnetic field in Au that exceeds the externally applied field, thus proving the existence of an intrinsic paramagnetic Meissner effect arising from an odd-frequency superconducting state.**

*Corresponding author. Email: jjr33@cam.ac.uk

# I. INTRODUCTION

Below the superconducting transition of a conventional (*s*-wave) Bardeen-Cooper-Schrieffer (BCS) superconductor such as Nb, the electrons stabilise into Cooper pairs in a spin-singlet state meaning that the electrons of a pair have oppositely aligned spins. The screening supercurrent density (*J*) that is generated by a superconductor in response to a weak magnetic field is linearly proportional to the vector potential (*A*) via the density of pairs present $n_s$ (*J*$=-e^2 n_s A/mc$, where *c*, *e* and *m* are the speed of light, the electron charge and the electron rest mass, respectively). Consequently, the amplitude of the screening supercurrent density (*J*) is negative and a diamagnetic Meissner effect is observed [1].

The opposite effect – the attraction of magnetic flux – has also been observed in superconductors [2-5], but this paramagnetic Meissner effect is metastable and is due to inhomogeneities and is not, therefore, intrinsic to the superconductivity. An intrinsic paramagnetic Meissner state has been predicted in *s*-wave superconductors with broken time-reversal symmetry [6-9], as a result of an emergent unconventional odd-frequency superconducting state, which competes with conventional (even-frequency) superconductivity (see [6-9] and related theory in [10]).

At the surface of an *s*-wave superconductor proximity coupled to a magnetic metal, the exchange field of the magnet can induce an odd-frequency superconducting state in which the Cooper pairs are in a spin-triplet state with a density ($n_t$) that is a mixture of spin-zero and spin-one pair projections [11-13]. This means that *J* is dependent on the magnitude and sign of $n_s$-$n_t$ (i.e. *J*$=-e^2(n_s - n_t)A/mc$) [14] and so odd frequency triplets should act to reduce the screening current [6]. Since $n_s$ and $n_t$ have different decay envelopes in an exchange field, *J* should reverse in sign as a function of magnetic layer thickness when $n_t$ exceeds $n_s$ at which point the magnetic susceptibility is positive and an inverse - paramagnetic - Meissner effect [6-9] prevails.

Evidence for spin-triplet pairing has recently been demonstrated in experimental studies involving magnetically inhomogeneous superconductor/ferromagnet (S/F) hybrids, such as via transition temperature measurements of S/F1/F2 spin valves [15], long-ranged supercurrents in S/F/S Josephson junctions [16], and various spectroscopy measurements of F/S systems [17].

To investigate the Meissner effect in a superconductor-magnet system, we measure the depth profile of the local magnetic susceptibility of a Au(27.5nm)/Ho(4.5nm)/Nb(150nm) trilayer by low-energy muon spin spectroscopy (LE-µSR). The antiferromagnetic rare earth metal Ho breaks time-reversal symmetry of the pair correlations in Au and has a thickness that is comparable to the known coherence length for singlet pairs in Ho [18] to ensure pair transmission into Au. The Au layer is necessary since a Meissner state cannot be probed by muons directly in a magnetic material due to their rapid depolarization in a strong magnetic field. Here, we report the discovery of the paramagnetic Meissner effect in Au, which is found to be an intrinsic property of the odd-frequency superconducting state that is generated via the superconductor proximity effect.

## II. RESULTS

LE-µSR offers extreme sensitivity to magnetic fluctuations and spontaneous fields of less than 0.1 Gauss with a depth-resolved sensitivity of a few nanometers [19-23]. To probe the depth dependence [$z$ coordinate in Fig. 1(a)] of the Meissner response in Au/Ho/Nb by LE-µSR, an external field ($B_{ext}$) is applied parallel to the sample plane [along the $y$ coordinate in Fig. 1(a)] and perpendicular to the muon initial spin polarization (oriented in the $x$-$z$ plane) as sketched in Fig. 1(a). The muon spin polarization is proportional to the asymmetry of decay positrons from the implanted muons as shown in Fig. 1(b), which is experimentally determined as a difference in the number of counting events of the two detectors, as discussed in the Supplemental Material [24].

In this transverse-field (TF) configuration, a muon's spin polarization precesses on average at a frequency $\bar{\omega}_s = \gamma_\mu \bar{B}_{loc}$ about the average local field $\bar{B}_{loc}$ sensed by the implanted muons, with $\gamma_\mu = 2\pi * 135.5$ MHz T$^{-1}$ being the muon gyromagnetic ratio. Assuming a local field profile $B_{loc}(z)$ within the sample, muons implanted with energy $E$ and a corresponding stopping distribution $p(z,E)$, precess at an average frequency $\bar{\omega}_s = \gamma_\mu \bar{B}_{loc} = \gamma_\mu \int B_{loc}(z) p(z,E) dz$ [25]. The asymmetry spectrum $A_s(t,E)$ can be approximated as $\propto e^{-\bar{\lambda}t}$ cos[$\gamma_\mu \bar{B}_{loc} t + \varphi_0(E)$] for each implantation energy $E$ ($\bar{\lambda}$ and $\varphi_0(E)$ being the mean depolarization rate and starting phase of the muon precession, respectively). The experimental $B_{loc}(z)$ profile is therefore sampled as series of mean field values $\bar{B}_{loc}$ of the magnetic field distributions $p(B_{loc})$ [Fig. 1(c)] determined as fits of the corresponding asymmetry functions $A_s(t,E)$ measured at different energies $E$ [24].

To investigate the paramagnetic Meissner effect, implantation energies in the 3-6 keV range were used to determine the $B_{loc}(z)$ profile in the Au layer. At the lowest energy of 3 keV, the muons contributing to the asymmetry stop within the Au, while for increasing energy, an increasing fraction stops within the Ho and Nb layers [Fig. 1(a) and Fig. S1 in the Supplemental Material [24]]. The implantation profiles were calculated using the Monte Carlo algorithm TrimSP [26]. To minimize the contribution from backscattered muons [26] to the measured signal, implantation energies below 3 keV were not used. For muon energies above ~7 keV, the contribution of the Nb becomes dominant and therefore not relevant for probing the Meissner state in the Au (Fig. S1 [24]). However, energies above 7 keV are important to confirm the emergence of a conventional (diamagnetic) Meissner response in Nb in the superconducting state. Muons stopping in the Ho layer depolarize almost immediately and do not contribute to the measured asymmetry.

Figures 2(a) and 2(b) show the $\bar{B}_{loc}$ values as function of implantation energy, obtained from fits to the data in the normal state at $T = 10$ K and in the superconducting state at $T = 5$ K (the critical temperature $T_c$ is ~ 8.52 K for the multilayer as reported in Fig. S2 of the Supplemental Material [24]). Figures 2(c) and 2(d) show similar plots but following warming and re-cooling with data taken at $T = 10$ K and $T = 3$ K. 6 x 10$^6$ positron-counting events were collected per datum point. The average stopping depth $\bar{z}(E)$ of the muon stopping profiles $p(z,E)$ for the corresponding implantation energy $E$ are plotted on the top axes in Fig. 2. The non-linearity of these depth scales stems from the fact that $\bar{z}(E)$ does not increase proportionally with $E$, as shown in Fig. S1(b) of the Supplemental Material [24]. Since the two energy scans at 3 K and 5 K were performed at different stages, the normal-state (10 K) energy scan is reacquired to avoid any influence on the measurement data from the specific magnetic configuration reached by Ho after cooling through its magnetic transitions.

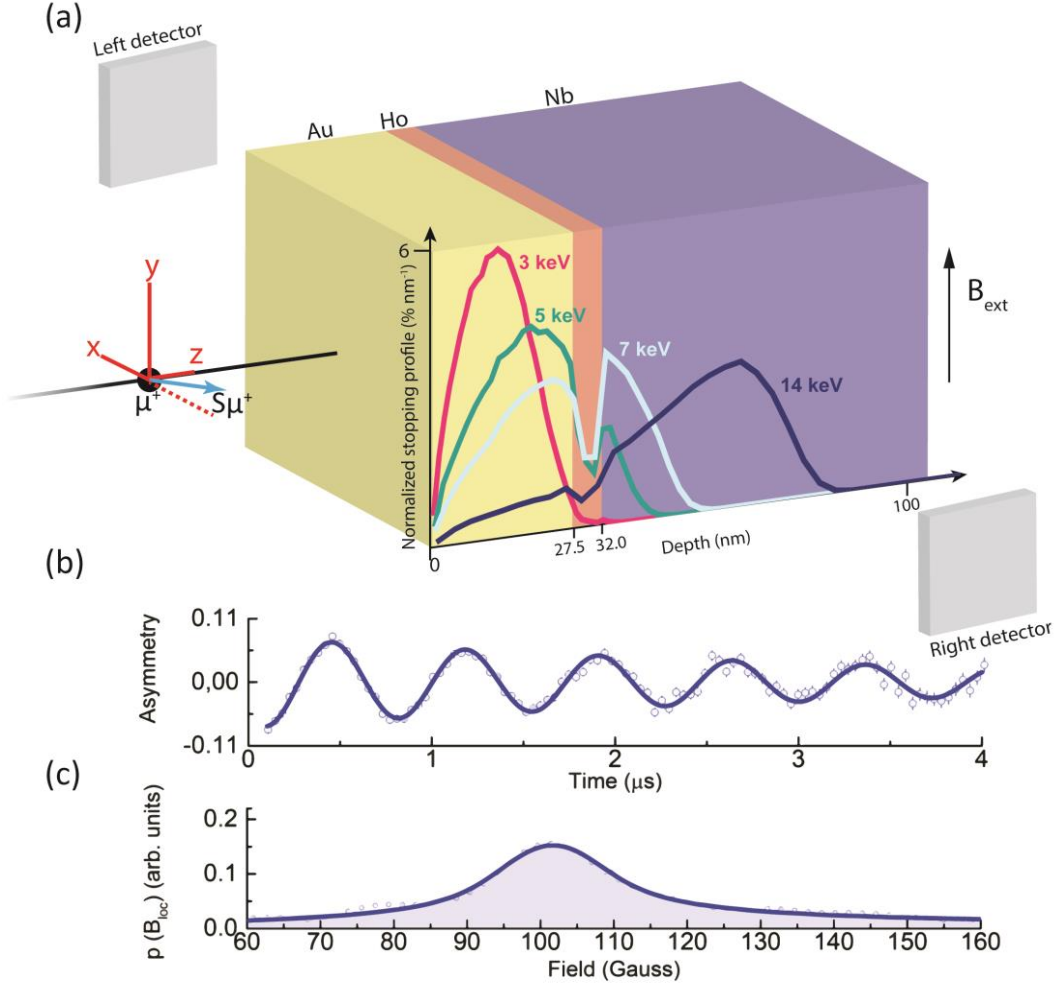

FIG. 1. Simulated muon stopping distributions in Au/Ho/Nb. (a) Experimental LE-µSR setup in transverse-field configuration and normalized muon stopping profiles $p(z,E)$ in Au/Ho/Nb simulated for a few representative implantation energies $E$. (b) Experimental asymmetry data determined from the counting events of the positrons detectors for muons implanted in Au/Ho/Nb with $E = 4.5$ keV at 3 K (blue dots) and single-energy asymmetry fit (blue curve). (c) Fourier transform of the asymmetry in (b) which represents the magnetic-field distribution.

The normal-state (10 K) data in Fig. 2 show that $\bar{B}_{loc}$ is approximately depth-independent and closely matches the externally applied field value $B_{ext}$ of about 101 Gauss [magenta curves in Figs. 2(a) and 2(c)]. The superconducting-state data (5 and 3 K), however, show a more complex behaviour. In Nb, the local field decreases as a function of depth in Nb and, for both temperatures, reaches a flux expulsion of about 2.5 Gauss for the 17 keV scan [corresponding to $\bar{z} \sim 29.7$ nm from the Ho/Nb interface as shown in Fig. S1(b) in the Supplemental Material [24]], consistent with conventional diamagnetic Meissner screening. In contrast, the opposite behaviour is observed in Au. Here, the local field increases by about 0.5 Gauss at $T = 3$ K and about 0.25 Gauss at $T = 5$ K above the applied field (and therefore the normal-state value of $B_{loc}(z)$) indicating a paramagnetic screening where $B_{loc}(z)$ appears non-monotonic with depth [blue curves in Fig. 2(a) and Fig. 2(c)]. We note that comparative LE-µSR studies by Morenzoni *et al.* on normal metal/Nb (N/Nb) bilayers demonstrate a purely diamagnetic response in the normal metal in the superconducting state, as expected due to the absence of a magnetic interface [27]. Although the measured increase in the local field, $\Delta B_{loc}(z)$, in the superconducting state relative to the normal state is small, it exceeds the statistical and systematic measurement error in $B_{loc}(z)$ (error bars in Fig. 2). Furthermore, $\Delta B_{loc}(z)$ increases at lower temperatures, which implies that the magnitude of $\Delta B_{loc}(z)$ is related to the amplitude of the superconducting order parameter which is consistent with theory [7,8].

Although the asymmetry fits for single implantation energy in Fig. 2 show a paramagnetic Meissner effect in Au below the Nb superconducting transition, the $B_{loc}(z)$ profiles obtained with this approach include depth averaging due to the width of muon stopping distributions. To obtain an accurate $B_{loc}(z)$ profile, a global fit for all implantation energies with a common field profile is used [19-23]. The common field profile in the Au layer is modelled as $B_{loc}(z) = B_{ext} + M(z)$, where we set the magnetization term to $M(z) = B_a \sin(z/\kappa)$, which is a parameterization of the theoretical magnetization profile calculated for the Au/Ho/Nb heterostructure.

The theoretical magnetization is computed from the vector potential *A* determined as a solution of the Maxwell equation $\frac{d^2 A}{dz^2} = -J = -J_x(z)A$, where the supercurrent *J* is assumed proportional to the vector potential *A* via the term $J_x(z)$ including the dependence on the anomalous Green's function (see Supplemental Material [24]). In this expression, $J_x(z)$ also represents the component of the supercurrent density *J* along the *x*-axis in Fig. 1(a). Both odd-frequency and even-frequency pairing correlations contribute to *J*, which is calculated using the quasiclassical theory of superconductivity under the assumption that time-reversal symmetry is spontaneously broken by the spatially-dependent exchange field of the Ho which forms a conical pattern along the *z* coordinate in Fig. 1(a). We also take into account the spin-selective scattering taking place at the interface between Nb and Ho by using spin-dependent boundary conditions [24]. Our model excludes the presence of a Fulde-Ferrel-Larkin-Ovchinnokov (FFLO) state which can theoretically compete with the paramagnetic Meissner state, but only if the superconducting layer is thinner than the magnetic screening length [6]. In Nb the magnetic screening length is about 90 nm, which is much shorter than the thickness of the Nb used here of 150 nm, and so contributions from the FFLO state can be ignored, as stated in Ref. [6].

In a normal metal (N) proximity-coupled to a superconductor (S), only even-frequency pairing correlations contribute to the screening supercurrent induced in N. The theoretical $M(z)$ profile in this case is $\propto \cosh(kz)/\cosh(k)$ (*k* being a measure for the supercurrent magnitude depending on several parameters including the thickness of the S/N bilayer, the diffusion constant*s* and the superconducting gap; for details see Supplemental Material [24]), which represents a monotonic decay from the S/N interface as expected for a conventional (diamagnetic) Meissner effect. In the presence of additional odd-frequency pairing correlations in the screening supercurrent induced by a magnetically-active layer separating the S/N interface (Ho in our case) instead, $B_{loc}(z)$ in N shows an oscillatory behaviour about $B_{ext}$ assuming both positive and negative values. In the particular case of Au/Ho/Nb, using realistic values for the physical parameters involved in the description of the proximity effect occurring in Au and Ho, the expected theoretical profile for $B_{loc}(z)$ shows a single oscillation reaching a maximum inside Au (blue curve in Fig. 3). Therefore, making also the realistic assumption that $B_{loc}(z)$ matches the applied external field $B_{ext}$ at the Au/vacuum interface, it is clear that $B_{loc}(z) = B_{ext} + B_a \sin(z/\kappa)$ represents an appropriate parameterization of the oscillatory local magnetic field profile in Au to use in the global energy fit. This parametrization is also in agreement with the experimental profiles determined at 3 and 5 K by sampling $B_{loc}(z)$ at different energies, which can be approximated by half-period sine functions (blue curves for Au in Fig. 2).

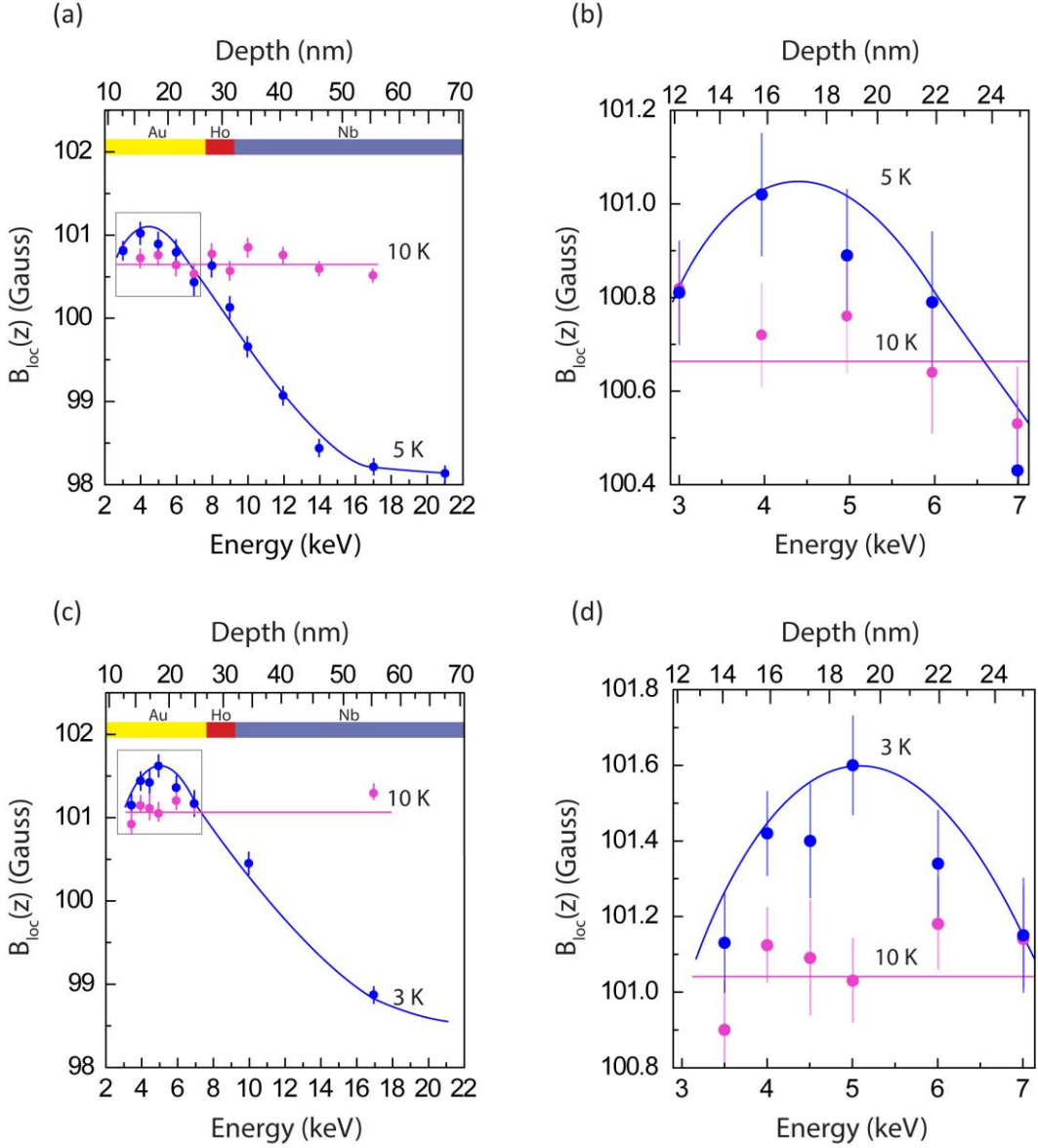

FIG. 2. Average local magnetic field in Au/Ho/Nb as a function of the muon implantation energy and mean stopping distance. (a) $\bar{B}_{loc}$ values from single-energy asymmetry fits versus implantation energy (bottom *x*-axis) and mean stopping distance (top *x*-axis) in the normal state (magenta circles 10 K) and superconducting state (blue circles 5 K), and in (b) identical data showing the inverse Meissner state in Au. The continuous lines are a guide to the eye. (c) and (d) Re-measured data after warming and cooling but the superconducting state is now measured at 3 K.

The global energy fit was implemented on the measurement data at 3 K which show the most significant paramagnetic Meissner response in Au. An exponential depolarization function $G(t,E) = e^{-\bar{\lambda}t}$ was used for the fit, with $\bar{\lambda}$ being a fitting parameter common for all energies [24]. In the analytic expression for $B_{loc}(z)$, $B_{ext}$ was set equal to the normal-state field obtained from the global fit of the measurement data at 10 K under the assumption that $B_{loc}(z)$ can be modelled as a constant field at this temperature [magenta curve in Fig. 2(c)]. Figure 3 illustrates the results of the global fit at 3 K which verify a positive increase of $B_{loc}(z)$ in Au over $B_{ext}$. The chi-square minimization algorithm reported good convergence (chi-square/numbers of degrees of freedom = 1.072) yielding $B_a$ = 0.55 Gauss and $\bar{\lambda}$ = 0.229 μs$^{-1}$ as optimal fitting parameters. The parameter $\kappa$ was kept fixed and equal to 13.58 nm to match the position of the peak in the $B_{loc}(z)$ theoretical profile.

The $B_{loc}(z)$ profile obtained with these values for $\bar{\lambda}$, $B_a$ and $\kappa$ (red curve in Fig. 3) is in good agreement with the theoretical $B_{loc}(z)$ curve (blue curve in Fig. 3) thus validating the good convergence of the algorithm and the appropriateness of the parameteric model used for the fitting. In addition, even when $\kappa$ was allowed to vary, the fits converged to a $\kappa$ value that is not significantly different (~15.2 nm), while $B_a$ remained the same, attesting further to the appropriateness of the model.

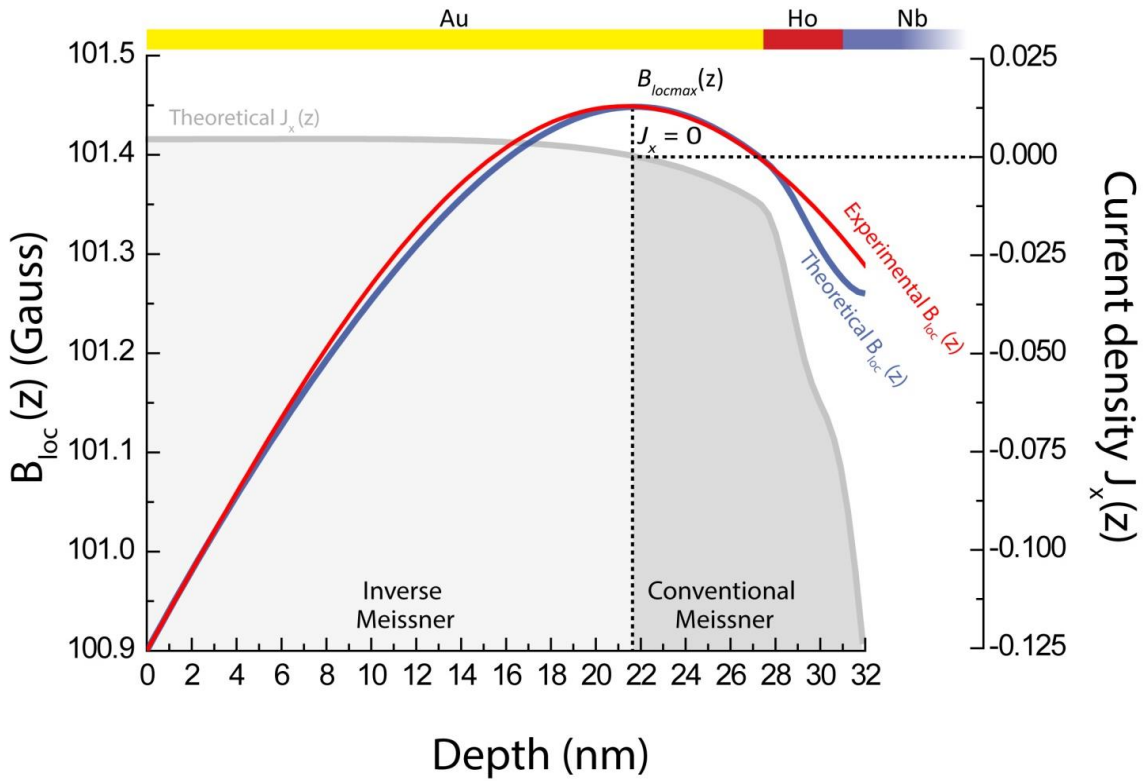

FIG. 3. Magnetization and screening current profile from the Ho/Nb interface in Au/Ho/Nb. Local magnetic field $B_{loc}(z)$ determined as global-energy fit of the LE-μSR measurement data (red curve, left y-axis) and theoretical model (blue curve, left y-axis); calculated dimensionless screening current density $J_x(z)$ flowing parallel to the x-axis in Fig. 1(a) inside the plane of the thin film heterostructure (grey curve, right y-axis). Dashed lines show that the position of the maximum in $B_{loc}(z)$ coincides with that of the null in $J_x(z)$.

# III. DISCUSSION

While the fits to the data are in close agreement with the predicted paramagnetic Meissner effect, we first rule out other possibilities that could lead to an increase in magnetic flux in Au. However, as we discuss below, none of these should show temperature dependence in the measured range of 3-10 K. One mechanism that can result in a magnetization enhancement in Au is Ruderman-Kittel-Kasuya-Yosida (RKKY)-type oscillations in the spin-polarization of Au induced via an interaction with Ho. The largest period predicted [28] and reported experimentally [29] for RKKY oscillations in epitaxial Au (001) is 7-8 monolayers or ~ 1.6 nm, which is much too small to explain the behaviour observed in Fig. 2, where the oscillation period of the magnetization exceeds several tens of nanometers. Furthermore, RKKY oscillations should also lead to an additional broadening of the magnetic field distribution experienced by muons (other than to the measured shift in average field) [30] and also be present in the normal-state data at 10 K, which we do not observe. A second possibility is an enhancement of the magnetization of Ho for decreasing temperature. Similarly to the case of RKKY oscillations, however, such an enhancement should only result in a broadening of the magnetic field distribution rather than the observed shift because the magnetic domains have a finite size and a random orientation giving a random dipolar field profile in Au.

Oscillations in the magnetic susceptibility induced by unconventional (odd-frequency) superconductivity are therefore the most likely explanation for the paramagnetic Meissner effect in Au due to the presence of Ho [6-9]. In Fig. 2 it is shown that a conventional Meissner effect is measured in Nb up to the interface with Ho, where the contribution of spin-singlet Cooper pairs to the screening supercurrent $J_x(z)$ is larger than that due to the long-ranged spin-triplet pairs (Fig. 3). Nevertheless, while spin-singlet pairs are rapidly filtered out by the exchange field in Ho, the spin-triplet pairs rotate into each other within the Ho layer (Fig. S3(a) in the Supplemental Material [24]) so that the sum of the Green's function amplitudes for spin-one and spin-zero triplet pairs follows a much slower decay in Ho compared to the Green's function amplitude for spin-singlets [Fig. S3(b)]. This is also consistent with the trend of the density of the screening current $J_x(z)$ profile (Fig. S3(b) in the Supplemental Material [24]), which depends on the imaginary part of the anomalous Green's functions via the difference $n_s$-$n_t$ between spin-singlet and spin-triplet pair densities: $J_x(z)$ starts off negative in Nb (conventional Meissner state), then increases inside Ho and it eventually becomes positive in Au, where the total contribution from the spin-triplets to the screening current overtakes the singlet one.

Finally, the results in Fig. 3 demonstrate that the paramagnetic effect is most strong in Au where the odd frequency state dominates over the singlet state. This indicates that the paramagnetic response is an intrinsic property of the odd frequency superconducting state and that the superconductivity must therefore carry a net magnetization. Future experiments should explore ways to harness the magnetization generated by odd frequency superconductivity in order to explore the potential for driving magnetization-reversal process in the superconducting state [13].

# ACKNOWLEDGMENTS

The authors thank N. Banerjee for sample growth support during the initial stages of the project. J.W.A.R. acknowledges financial support from the Royal Society through a University Research Fellowship. J.W.A.R. and A.D.B. acknowledge financial support from the UK EPSRC through NanoDTC EP/G037221/1 and the Leverhulme Trust through an International Network Grant (IN-2013-033). A.D.B. also acknowledges additional financial support from the Schiff Foundation. X.L.W. and J.H.Z. acknowledge support from the MOST of China (2015CB921500). J.L. acknowledges support from the Outstanding Academic Fellows programme at NTNU, the Norwegian Research Council Grant (205591, FRINAT, 216700). J. L., J.W.A.R, and A.D.B. finally acknowledge support from the COST Action MP-1201 'Novel Functionalities through Optimized Confinement of Condensate and Fields.' S.L. and M.G.F. acknowledge the support of the EPSRC through Grant No. EP/J01060X. The muSR measurements were performed at the Swiss Muon Source (SµS), at the Paul Scherrer Institute in Villigen, Switzerland. The project has also received funding from the European Union's Seventh Framework Programme for research, technological development and demonstration under the NMI3-II Grant number 283883. Data accompanying this publication are directly available at the https://www.repository.cam.ac.uk/handle/1810/251382 data repository.

# Supplemental Material

## I. SAMPLES PREPARATION AND LE-µSR MEASUREMENT SETUP

The Au(27.5nm)/Ho(4.5nm)/Nb(150 nm) thin film multilayers were grown onto unheated oxidised Si substrates by DC magnetron sputtering in Ar plasma at 1.5 Pa. Before and during the film growth, the walls of the deposition chamber were cooled via a liquid nitrogen jacket to achieve a base pressure of below $10^{-8}$ Pa (verified using an in-situ residual gas analyser). The substrates were placed on a circular table and rotated below stationary sputtering targets. Deposition rates were pre-determined for each material using an atomic force microscope to measure pre-deposited step edges on calibration samples. The calibrated deposition rates were used to set the rotation speeds and deposition powers needed to achieve the desired thickness for each metallic layer.

In contrast to bulk µSR where ~4 MeV muon beams are used, in the LE-µSR apparatus at Paul Scherrer Institute (PSI) lower muon implantation energies can be obtained using a few-hundred-nanometer-thick Ar solid gas moderator capped with ~ 10 nm $N_2$ and grown on top of a silver foil [31] (about 100 µm in thickness). Choosing appropriate beam transport and sample voltages, the muon implantation energy can be varied between 0.5 keV and 30 keV, which allows tuning of the mean muon stopping depth in the range 1-200 nm with an accuracy of a few nanometers. The energies needed to probe the local magnetic field at specific depths inside the Au/Ho/Nb samples were chosen on the basis of the stopping profiles calculated using a Monte Carlo algorithm TRIM.SP [32]. A transverse-field geometry [33] was adopted with the external field $B_{ext}$ ~ 101 Gauss applied in the film plane and perpendicular to the initial polarization of the muons in order to bring the sample into a Meissner state. This configuration, usually adopted when $B_{ext}$ is larger than any internal field, is extremely sensitive to sample magnetic inhomogeneities [34], which would result in a spreading of the Larmor frequencies for the individual muon decay events. LE-µSR measurements were carried out at two different temperatures (3 K and 5 K) below the Nb superconducting transition, with 3 K being the lowest temperature achievable with the measurement apparatus at PSI.

## II. THEORY AND FITTING OF THE LE- µSR MEASUREMENTS

A schematic diagram of the apparatus used to probe the local magnetic field of our Au/Ho/Nb heterostructure by LE-µSR has been reported in Fig. 1(a). In a transverse-field (TF) configuration, where the initial polarization of the muon spin is perpendicular to the applied field, the starting point for the analysis of the LE-µSR measurement data is given by the counting events recorded by the left and right positron detectors $N(t,E)$ as a function of time $t$ and muon implantation energy $E$:

$$N_L(t, E) = N_0 e^{-\frac{t}{t_\mu}}[1 + A_s(t, E)] + K_L \quad (1.1)$$

$$N_R(t, E) = \alpha_d N_0 e^{-\frac{t}{t_\mu}}[1 - A_s(t, E)] + K_R \quad (1.2)$$

where the subscripts *L* and *R* denote the left and right positron detector respectively, *K* is the time-independent background contribution due to accidental coincidences, $t_\mu \sim 2.2$ µs is the muon lifetime, $\alpha_d \sim 1$ is the detector efficiency correction factor and $A_s(t,E)$ is the asymmetry function which is proportional to the muon spin polarization.

From equations (1.1) and (1.2) we can calculate $A_s(t,E)$ as the difference between the counting events of the left and right positron detectors normalized by their sum:

$$A_s(t,E) = \frac{\alpha_d[N_L(t,E)-K_L]-[N_R(t,E)-K_R]}{\alpha_d[N_L(t,E)-K_L]+[N_R(t,E)-K_R]} \tag{1.3}$$

The asymmetry function contains information on the spatial and temporal variation of the sample local magnetization $M(z)$. Making the realistic assumption that $B_{ext} \gg M(z)$ in the sample local magnetic field $B_{loc}(z)$ (where $B_{loc}(z) = B_{ext} + M(z)$) and that the local spin environment appears to the muons as static (i.e. the spin fluctuation frequencies are much lower than the inverse muon lifetime $t_\mu^{-1}$), $A_s(t,E)$ can be in the simplest form written as:

$$A_s(t,E) = A_{s0} \cos\left(\gamma_\mu \bar{B}_{loc} t + \varphi_o(E)\right) G(t,E) \tag{1.4}$$

with $\gamma_\mu = 2\pi*135.5$ MHz T$^{-1}$ being the muon gyromagnetic ratio, $A_{s0}$ the initial asymmetry that can be detected with the measurement setup (about 0.25), $\varphi_0(E)$ the starting phase of the muon precession, and $G(t,E) \leq 1$ the depolarization function due to inhomogeneities and/or dynamics in the local magnetic field. The expression (1.4) takes into consideration the stopping distribution $p(z,E)$ of the implanted muons at energy *E*, since $\bar{B}_{loc}$ is the weighted average of the local field $B_{loc}(z)$ over $p(z,E)$.

Equations (1.3) and (1.4) can be used in combination to perform an asymmetry fit at a specific energy *E*. For the system investigated, the best fits were obtained using an exponential depolarization function $G(t,E) = e^{-\bar{\lambda}t}$ ($\bar{\lambda}(E)$ being the mean depolarization rate at the energy *E*) rather than a more conventional Gaussian function, which can be ascribed to the presence of stray fields due to the Ho layer. The collection of the mean field values $\bar{B}_{loc}$ obtained at different energies with this fit provides a preliminary profile $B_{loc}(z)$ of the spatial distribution of the magnetic field in the Au/Ho/Nb sample. The depth within the multilayer corresponding to the $\bar{B}_{loc}$ value at energy *E* is set equal to the average stopping distance $\bar{z}(E)$ of the muon beam implanted with the same energy *E* [see Fig. S1(b)].

Significant improvement to the fit can be made by taking the contribution of the muon stopping profile $p(E,z)$ and the magnetic field profile into consideration when calculating the asymmetry signal. In order to perform this improved fit, the following more general expression is used for $A_s(t,E)$ instead of equation (1.4):

$$A_s(t,E) = \int p(E,z) A_{s0} \cos\left(\gamma_\mu B_{loc}(z) t + \varphi_o(E)\right) G(t,E) dz \tag{1.5}$$

where the integral is now extended over the entire sample depth range probed by the muon beam at a given implantation energy *E*. The local magnetic field $B_{loc}(z)$ is not treated as a constant field for a given muon implantation energy like in (1.4), and the goal is to determine a functional form for $B_{loc}(z) = B_{ext} + M(z)$ which is consistent with the measurement data simultaneously for all the energies considered. A good convergence of the chi-square minimization algorithm run by the software musrfit [35] (*chi-square/number of degrees of freedom* ratio close to 1) for all the energies is normally only achieved for an appropriate choice of the functional form for $B_{loc}(z)$.

Since the theoretical profile calculated for $B_{loc}(z)$ in Au can be parameterized by a half-cycle sinusoid with $B_{loc}(z) = B_{ext}$ at the Au/vacuum interface (blue curve in Fig. 3), we have used the following functional form for $B_{loc}(z)$ in (1.5) to perform the global energy fit:

$$B_{loc}(z) = B_{ext} + M(z) = B_{ext} + B_a \sin(z/\kappa) \qquad (1.6)$$

where the amplitude $B_a$ and the angular frequency $\kappa^{-1}$ of the sinusoid are the model parameters. This common field profile is also consistent with the trend of the experimental $\bar{B}_{loc}(z)$ values determined from single-energy asymmetry fits in the superconducting state at 3 K and 5 K (blue curves in Fig. 2).

Since the expression (1.6) does not apply to the local field profile in the Nb layer, only energies up to 6 keV were taken into account to implement the global fit. This is consistent with the simulated muon stopping fractions in Fig. S1(c), which show that the contribution of muons stopping in Nb to the asymmetry signal becomes non-negligible at energies higher than 6 keV. $B_{ext}$ in (1.6) was determined from the global fit of the measurement data in the normal state at 10 K assuming that $B_{loc}(z) = $ constant $= B_{ext}$ through the entire Au/Ho/Nb multilayer.

Using $\bar{\lambda}$ and $B_a$ as fitting parameters and fixing $\kappa = 13.58$ nm – which corresponds to a sine function having the same peak position inside Au as the theoretical $B_{loc}(z)$ profile (blue curve in Fig. 3) – the minimization algorithm converged with *chi-square/number of degrees of freedom* = 1.072 and it yielded $B_a = 0.549$ Gauss and $\bar{\lambda} = 0.229$ μs$^{-1}$.

When $\kappa$ was kept as free fitting parameter and not fixed to the theoretical value of 13.58 nm, $\kappa$ converged to a value slightly different (~15.2 nm), while *chi-square/number of degrees of freedom* and $B_a$ remained the same (1.072 and 0.55 Gauss, respectively). This result shows that the global fit of the experimental data tend to converge to the values predicted by theory independently on it.

## III.  ANALYTICAL DESCRIPTION OF THE ANOMALOUS MEISSNER RESPONSE

Normalizing the z-axis coordinate by $L=d_N+d_F$, where $d_N(d_F)$ is the thickness of the N(F) layer, so that $z=0$ corresponds to the Nb/Ho interface and $z=1$ to the Au/vacuum interface, the Maxwell equation that determines the magnetization response reads:

$$\frac{d^2 \mathbf{A}}{dz^2} = -\mathbf{J} = -J_x(z)\mathbf{A} \qquad (1.7)$$

where $\mathbf{A}$ is the vector potential and $J_x(z)$ is the normalized screening current density along the x-axis in Fig. 1. Using the linear response theory formalism, the screening supercurrent density $\mathbf{J}$ of Maxwell equation has been written in equation (1.7) as proportional to the vector potential **A** via a factor $J_x(z)$ containing the dependence on the anomalous Green's functions [36].

The induced magnetization **M** in the N layer and the vector potential **A** are related through the following expression:

$$\boldsymbol{M} = \frac{dA}{dz} - 1 \qquad (1.8)$$

Here, we have normalized **M** against the external field $B_{ext}$, while the amplitude of $dA/dz$ is equal to that of the total local magnetic field $B_{loc}(z) = B_{ext} + M(z)$ normalized against $B_{ext}$. In our case, we use the following boundary conditions to solve equation (1.7) and then compute **M** from equation (1.8):

$$\boldsymbol{A}(0) = 0 \qquad (1.9.1)$$

$$\left.\frac{dA}{dz}\right|_{z=1} = 1 \qquad (1.9.2)$$

which mean, from a physical point of view, that the superconductor shields completely the external magnetic field and that there is no induced magnetization at the Au/vacuum interface. The equation (1.7) with the boundary conditions given by equations (1.9.1) and (1.9.2) is normally difficult to solve analytically [37]. Nevertheless, to understand the basic physics underlying the anomalous Meissner response using this equation, it is possible to consider some limiting cases.

Firstly, it can be proven [38] that the sign of the screening current in (1.7) is negative (positive) for a pure even (odd) frequency pairing in the N layer. For the case of a conventional Meissner response (even-frequency pairing only) and assuming for simplicity that the spatial dependence of $J_x(z)$ is negligible, equation (1.7) can be written as [38]:

$$\frac{d^2A}{dz^2} = k^2 \boldsymbol{A} \qquad (1.10)$$

where $k$ is a constant. Using the boundary conditions, it can be shown that equation (1.10) has the following solution:

$$M(z) = \frac{\cosh(kz)}{\cosh(k)} - 1 \qquad (1.11)$$

Since $z \in [0, 1]$, $M(z)$ in equation (1.11) is always negative and it decays monotonically from the N/F interface as expected for a conventional Meissner response. Similarly, if there is a constant odd-frequency supercurrent ($|J| > 0$), then equation (1.7) can be written as:

$$\frac{d^2A}{dz^2} = -k^2 \boldsymbol{A} \qquad (1.12)$$

which can also be solved analytically giving the following solution:

$$M(z) = \frac{\cos(kz)}{\cos(k)} - k \qquad (1.13)$$

The proximity-induced magnetization in equation (1.13) shows an oscillatory behaviour and it assumes both positive and negative values. This means that the induction of a shielding-supercurrent generated by odd-frequency pairing correlations does not necessarily give only an inverse (paramagnetic) Meissner response.

For our Au/Ho/Nb heterostructure, we expect that even- and odd-frequency pairings are mixed together in Au. Therefore, any signature of an anomalous Meissner effect in N implies that the odd-frequency component is dominant herein [38].

Equation (1.7) has been solved numerically for our Au/Ho/Nb system giving the $J_x(z)$ profile reported in Fig. 3. In this case, there is a point $z_0 \in [0, 1]$ where $J_x(z)$ changes sign and the contribution of odd-frequency correlations becomes dominant over the even-frequency one. Since $J_x(z)$ is the spatial derivative of $M(z)$ (by virtue of equations (1.7) and (1.8)), the observed peak in $M(z)$ at $z = z_0$ (Fig. 3) must be a signature of dominant odd-frequency pairing correlations in the N layer.

## IV. SOLUTION OF THE USADEL AND MAXWELL EQUATIONS

To quantify the proximity effect that generates superconducting correlations in the Ho and Au layers, we use the quasiclassical theory of superconductivity. The correlations are then described by a Green's function matrix $g$ that satisfies the Usadel equation in the non-superconducting region:

$$D\partial_z(g\partial_z g) + i[E_q\rho_3 + R(z), g] = 0 \qquad (1.14)$$

Here, $D$ is the diffusion constant, $E_q$ is the quasiparticle energy, $\rho_3 = \text{diag}(1,1,-1,-1)$, while $R(z)$ is a 4x4 matrix that contains the exchange field profile {for further details see [39]}. To compute $g$, one also needs to use boundary conditions at the interfaces of the junction. At the Au/vacuum interface, the derivative of the Green's function must be zero, while at the Nb/Ho interface we use the standard Kupryianov-Lukichev boundary conditions valid for a non-ideal interface (as this is experimentally realistic):

$$2\varsigma L g \partial_z g = [g_{BCS}, g] + iG_\phi[\beta, g] \qquad (1.15)$$

where $\varsigma$ is the ratio between the interface resistance and the resistance of the non-superconducting layer, $L = d_N + d_F$ is the length of the non-superconducting layer, $g_{BCS}$ is the bulk superconducting Green's functions, $G_\phi$ is a phenomenological parameter capturing the spin-dependent phase-shifts occurring at the interface, while $\beta$ is a matrix describing the orientation of the interface magnetic moment [39].

In the superconductor, we use the bulk superconducting Green's function since the superconductor is much larger than the other regions and it acts as a reservoir. To model the Ho/Au region, we use a spatially-dependent exchange field with a magnitude constant in Ho (rotating in direction) but gradually dropping to zero over a region of few nanometers centred at the Ho/Au interface. As long as this drop is not too sharp (i.e. occurring over 1 nm or less), this assumption is allowed by the quasiclassical formalism, which demands that all length-scales involved must be much larger than the Fermi wavelength (typically of the order of 1 Å).

Within the linear response theory, the amplitude $J$ of the shielding current density $\boldsymbol{J}$ generated in the junction reads:

$$J = \frac{-N_0|e|D}{16} \int Tr\{\rho_3(\breve{g}i|e|A[\rho_3,\breve{g}]_-)^K\}d\varepsilon \tag{1.16}$$

where $N_0$ is the normal-state density of states at the Fermi level, $e$ is the electron charge, $\breve{g}$ is the 8x8 Green's function in Keldysh space, $D$ the diffusion constant and $A$ the magnetic vector potential. The superscript 'K' means that the Keldysh component of the matrix between the parentheses should be taken.

This current density is then used as input in the Maxwell equation (1.7) that determines the vector potential $\boldsymbol{A}$, from which one computes the magnetic field $B(z)$ and magnetization $M(z)$ amplitudes. Note that in equation (1.7) we have separated the vector potential $\boldsymbol{A}$ explicitly from the current. In the fit to the experimental data, we have used the following parameters: $H_{ex}/\Delta = 18$ ($\Delta$ being the amplitude of the superconducting order parameter), $T/T_c = 0.4$, $d_F = 4.5$ nm, $d_N = 27.5$ nm, coherence length $\xi_s = 30$ nm, spiral length in Ho $\lambda_s = 3.4$ nm, and $G_\phi = 1.2$ with an interface magnetic moment lying in the plane of the Nb/Ho interface. The strength of the current density is determined by the fitting parameter $k = \frac{L^2\mu_0 N_0 e^2 D\Delta}{8\hbar}$ which was set equal to 2.9 (here, $\mu_0$ the vacuum permeability and $\hbar$ the reduced Planck constant). This corresponds to a reasonable set of values for the parameters entering the expression for $k$ such as $D \sim 10^{-3}$ m$^2$/s, $\Delta \sim 2$ meV, and $N_0 \sim 10^{24}$/(cm$^3$ eV).

# ADDITIONAL REFERENCES

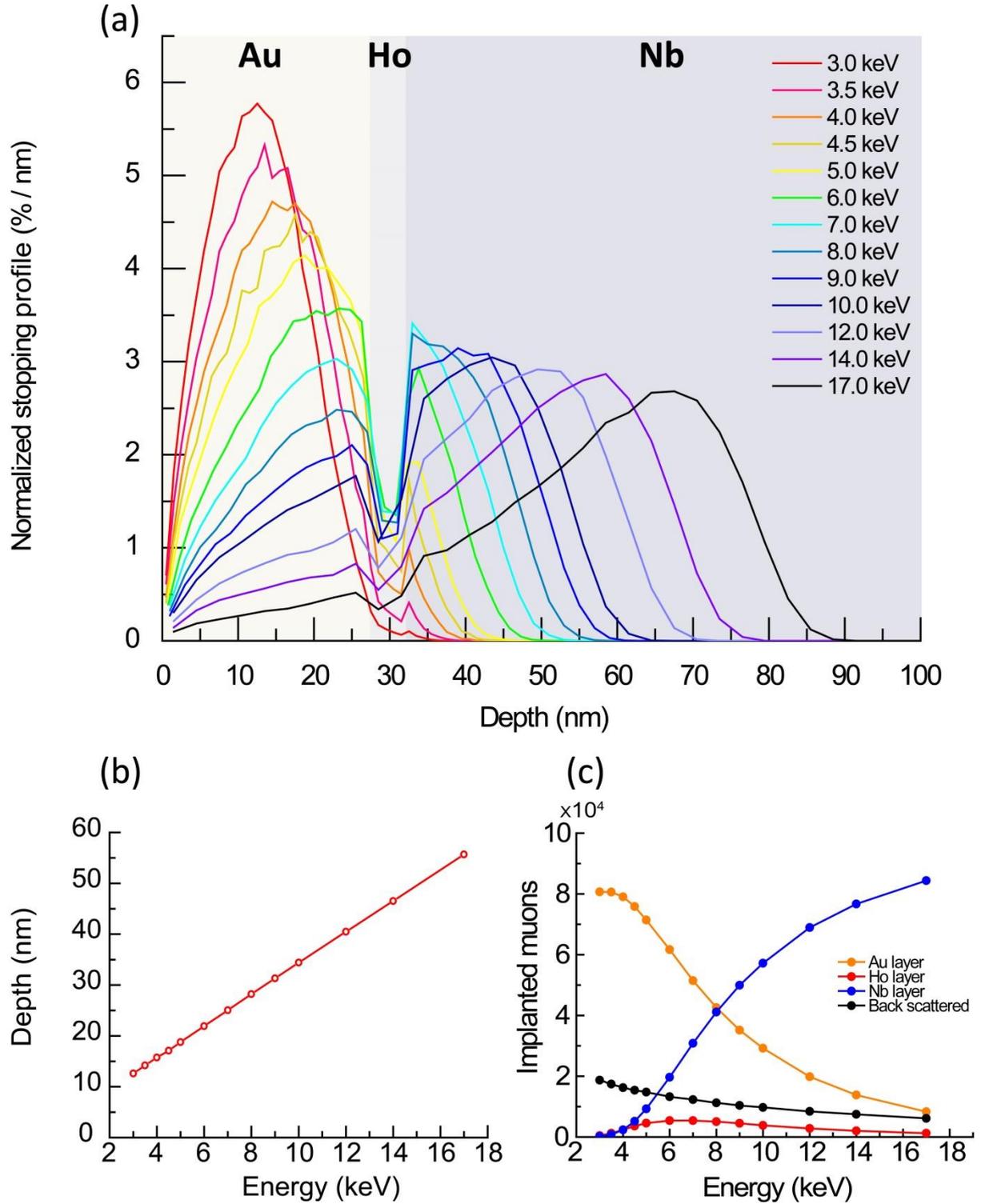

FIG. S1. Normalized stopping profiles, average stopping depth and fractions of implanted muons at different energies. (a) Simulated stopping profiles $p(z,E)$ for the Au(27.5 nm)/Ho(4.5 nm)/Nb(150 nm) heterostructure and in (b) corresponding average stopping depth $\bar{z}\,(E)$ as a function of the muon implantation energy $E$. The grey shaded areas in (a) define the limits of each layer within the heterostructure. (**c**), Simulated fractions of muon particles stopping in each layer of the Au(27.5 nm)/Ho(4.5 nm)/Nb(150 nm) heterostructure as a function of the muon implantation energy. The data in (c) show that for energies higher than 6 keV an increasingly significant number of muon particles probe the local magnetic field within the Ho and Nb layers.

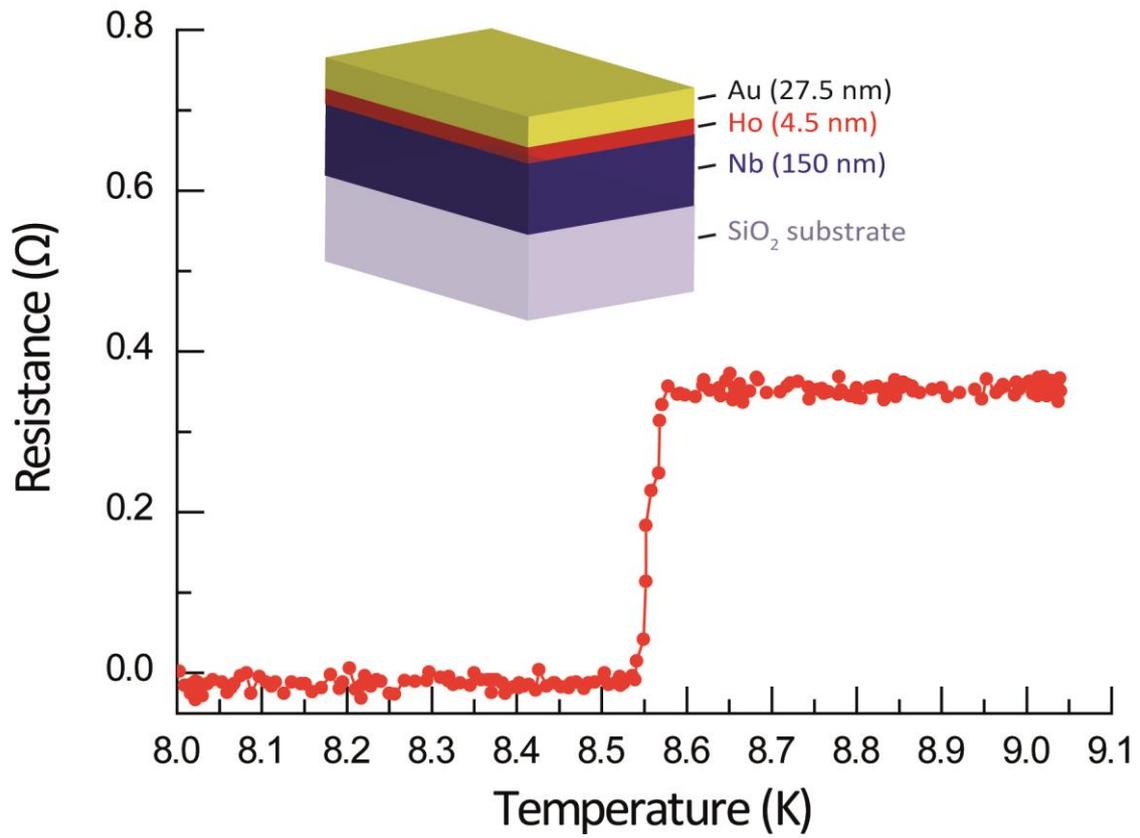

FIG. S2. Resistance versus temperature (R-T) plot for a Au(27.5 nm)/Ho(4.5 nm)/Nb (150 nm) film on $SiO_2$. Electrical transport measurements were performed in a four-point current-bias setup using a dipstick probe in liquid helium dewar. The superconducting transition temperature is roughly 8.52 K at 50% of the transition. During measurements it was ensured that the applied bias current (0.1 mA) had no effect on the $T_c$ value.

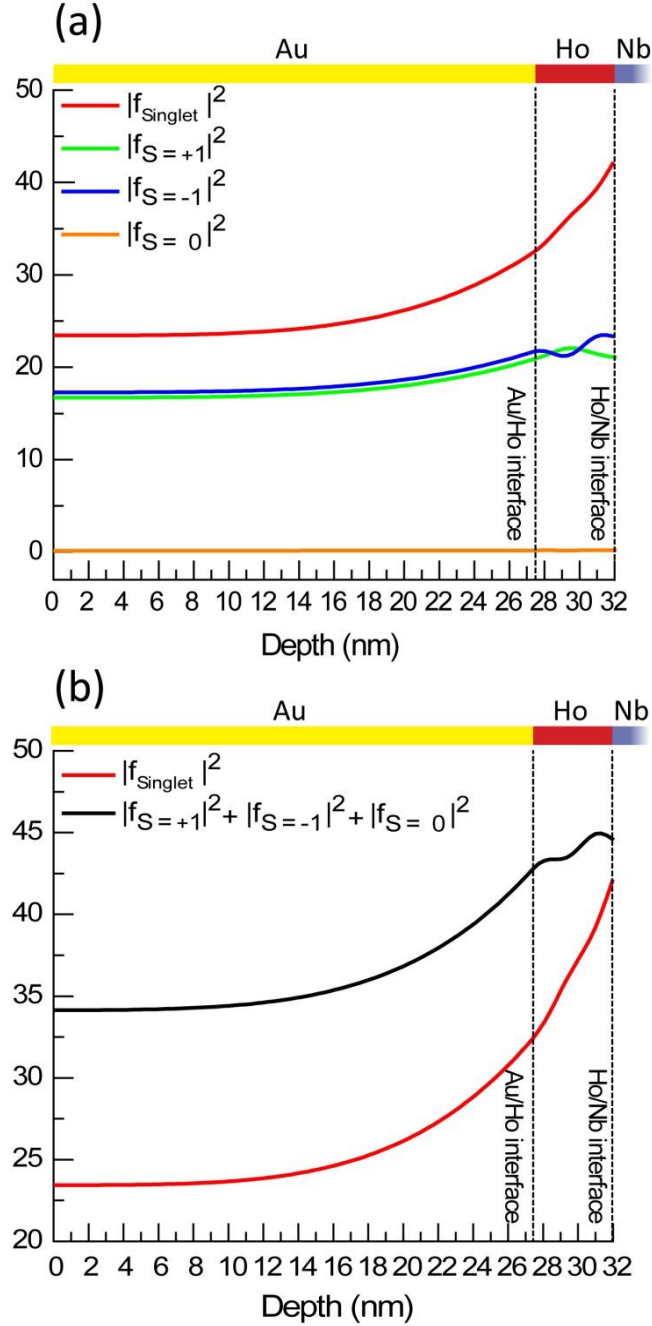

FIG. S3. Amplitude variation of the anomalous Green's functions for spin-singlet and spin-triplet Cooper pairs from the Ho/Nb interface. (a) Amplitude variation of the Green's function for spin-singlets ($f_{Singlet}$) and spin-triplets with spin-projection equal to +1 ($f_{S=+1}$), -1 ($f_{S=-1}$) and 0 ($f_{S=0}$). (b) Sum of the amplitudes of the spin-triplet terms versus the amplitude of the spin-singlet term. While fully-polarized spin-triplets ($S = \pm 1$) are rotated into each other by the Ho magnetic helix inducing an oscillatory behaviour of the corresponding function amplitudes in the Ho layer (green and blue curves), the amplitude of the singlet term (red curve) follows a fast decay.

The amplitudes of the anomalous Green's functions have been derived solving the exact Usadel equation in the non-linearized regime (absence of a weak proximity effect). Each anomalous Green's function amplitude is defined as sum over a fine mesh of energies between zero and ten times the superconducting gap. The quantization axis is taken to be along the interface magnetic moment. This figure is meant to illustrate the different behaviour of the spatial evolution of the triplet and singlet components by defining $|f|^2$ as a measure of the strength of the proximity effect. We emphasize that even if the total triplet amplitude defined in this way is larger than the singlet even near the Nb interface, this does not mean that the triplet contribution to the actual screening current necessarily is larger in this region, since this quantity depends on the anomalous Green's functions in a different way than $|f|^2$.